%% file: dipole.tex
\documentclass[twocolumn]{aa}

\usepackage{dsfont}
\usepackage{graphicx}
\usepackage{amsmath,amsfonts,amssymb}
\usepackage{txfonts}
\usepackage{color}
\usepackage{natbib}
\usepackage{float}
\usepackage{dblfloatfix}
\usepackage{afterpage}
\usepackage{ifthen}
\usepackage[morefloats=12]{morefloats}
\usepackage{placeins}
\usepackage{multicol}
\usepackage{subcaption}
\bibpunct{(}{)}{;}{a}{}{,}
\usepackage[switch]{lineno}
\definecolor{linkcolor}{rgb}{0.6,0,0}
\definecolor{citecolor}{rgb}{0,0,0.75}
\definecolor{urlcolor}{rgb}{0.12,0.46,0.7}
\usepackage[breaklinks, colorlinks, urlcolor=urlcolor,
    linkcolor=linkcolor,citecolor=citecolor,pdfencoding=auto]{hyperref}
\hypersetup{linktocpage}

\input{Planck}

\def\nside{N_{\mathrm{side}}}

\def\healpix{\texttt{HEALPix}}

\def\npipe{\texttt{NPIPE}}
\def\Planck{\textit{Planck}}

\renewcommand{\d}[0]{\vec{d}}

\newcommand{\x}[0]{\vec{x}}
\newcommand{\y}[0]{\vec{y}}
\newcommand{\Y}[0]{\tens{Y}}
\newcommand{\n}[0]{\vec{n}}

\newcommand{\s}[0]{\vec{s}}
\renewcommand{\a}[0]{\vec{a}}

\newcommand{\T}[0]{\tens{T}}

\renewcommand{\L}[0]{\tens{L}}

\newcommand{\N}[0]{\tens{N}}
\newcommand{\M}[0]{\tens{M}}

\renewcommand{\S}[0]{\tens{S}}

\def\inv{^{-1}}

\begin{document}

\title{A Monte Carlo comparison between template-based and Wiener-filter CMB dipole estimators}
\author{\small
  H.~Thommesen\inst{1}\thanks{Corresponding author: H.~Thommesen; \goodbreak \url{harald.thommesen@astro.uio.no}}
  \and
  K. J. ~Andersen\inst{1}
  \and
  R. ~Aurlien\inst{1}
  \and
  R. ~Banerji\inst{1}
  \and
  M. ~Brilenkov\inst{1}
  \and
  H. K. ~Eriksen\inst{1}
  \and
  U. ~Fuskeland\inst{1}
  \and\\
  M. ~Galloway\inst{1}
  \and
  L. M. ~Mocanu\inst{1}
  \and
  T. L. ~Svalheim\inst{1}
  \and
  I. K. ~Wehus\inst{1}
}
\institute{\small
  Institute of Theoretical Astrophysics, University of Oslo, P.O.Box
  1029 Blindern, N-0315 Oslo, Norway\goodbreak
}

\authorrunning{H. Thommesen et al.}
\titlerunning{A Monte Carlo comparison between template-based and Wiener-filter CMB dipole estimators}

\abstract{We review and compare two different CMB dipole estimators
  discussed in the literature, and assess their performances through
  Monte Carlo simulations. The first method amounts to simple template
  regression with partial sky data, while the second method is an
  optimal Wiener filter (or Gibbs sampling) implementation. The main
  difference between the two methods is that the latter approach takes
  into account correlations with higher-order CMB temperature
  fluctuations that arise from non-orthogonal spherical harmonics on
  an incomplete sky, which for recent CMB data sets (such as \Planck)
  is the dominant source of uncertainty. For an accepted sky fraction
  of 81\,\% and an angular CMB power spectrum corresponding to the
  best-fit \Planck\ 2018 $\Lambda$CDM model, we find that the
  uncertainty on the recovered dipole amplitude is about six times
  smaller for the Wiener filter approach than for the template
  approach, corresponding to 0.5 and 3\,\muK, respectively. Similar
  relative differences are found for the corresponding directional parameters and
  other sky fractions. We note that the Wiener filter algorithm is
  generally applicable to any dipole estimation problem on an
  incomplete sky, as long as a statistical and computationally tractable
  model is available for the unmasked higher-order fluctuations. The
  methodology described in this paper forms the numerical basis for
  the most recent determination of the CMB solar dipole from \Planck,
  as summarized by \citet{npipe}. }


\maketitle

\section{Introduction}
\label{sec:introduction}
The cosmic microwave background (CMB) radiation was discovered in 1965
by \citet{penzias:1965}, and has ever since been the primary target
for several dozens of CMB experiments. The main scientific target for
most of these studies has been small variations in intensity and
polarization that correspond to cosmic density variations some
380\,000 years after the Big Bang. These variations contain a wealth of
information about the early history and evolution of the universe; for
a recent analysis, see, e.g., \citet{planck2016-l06}.

The CMB sky features three main physical components. The first is
simply a constant blackbody term with a temperature of 2.7255\,K
\citep{fixsen:2009}, corresponding to the average temperature of the
CMB photons populating the universe today. This component is often
denoted the CMB monopole, acknowledging its correspondence to the
lowest multipole moment in spherical harmonics space.

The second component is the CMB dipole, which has an amplitude of
about 3\,mK \citep{lineweaverdipole}.  The CMB dipole is the result of
Doppler boosting caused by the motion of the measuring instrument with
respect to the CMB rest frame. It may be decomposed into two
components, namely the \textit{solar dipole} caused by the movement of
the solar system around the Milky Way's center, and the
\textit{orbital dipole} generated by the movement of the Earth and the
instrument around the Sun. Both components play an important role for
CMB experiments, as they represent the best available astrophysical
calibration source for most experiments. Specifically, the orbital
dipole serves as an invaluable tool for absolute calibration, since
the Earth--Sun distance and the orbital period are known with very high
precision. Likewise, the solar dipole provides an excellent relative
calibration target, since it is brighter than most other signals, has
a perfectly known frequency spectrum, and is visible across the entire
sky.

The third component is the CMB density fluctuations with typical
variations of about 100\muK. These correspond very closely to a
statistically isotropic and Gaussian random field with an angular
power spectrum described by a $\Lambda$CDM power spectrum
\citep{planck2016-l06,planck2016-l07}. In addition to these CMB
sources, real-world microwave observations also contain
contributions from astrophysical foregrounds, most notably in the form
of synchrotron, free-free, spinning and thermal dust, and CO emission
from the Milky Way \citep[e.g.,][]{planck2016-l04}.

This paper will discuss how to optimally estimate the amplitude and
direction of the CMB dipole with data that contains both astrophysical
foregrounds and small-scale CMB fluctuations. Obtaining robust
estimates for these parameters is important for several
reasons. First, since the CMB dipole is used as a calibration source
for most experiments, a potential bias in the CMB dipole amplitude
translates directly into a corresponding bias of the overall
normalization of the angular CMB power spectrum. Second, because the
CMB dipole is about four orders of magnitude brighter than the
cosmological variations in the large-scale polarization field, it is
necessary to estimate the relative gains between detectors within a
single frequency channel prior to mapmaking with a precision better
than $\mathcal{O}(10^{-4})$ in order to avoid significant bias on the
optical depth of reionization, $\tau$ \citep{planck2016-l06}.

Uncertainties on the dipole parameters result mainly from four
different contributors. First, statistical instrumental noise defines
a fundamental floor for the overall sensitivity that can be
achieved. Second, many systematic effects due to non-idealities in the
instrument itself can induce spurious dipoles, including sidelobe
pickup, time-variable gain, or ADC corrections. Third, as already
mentioned, foreground emission from the Milky Way obscures our view of
the CMB, and also carries a dipole moment. To mitigate this effect it is
in practice necessary to mask out parts of the Galactic
regions. However, working with incomplete sky coverage has the
unwanted effect of making the spherical harmonic base functions,
$Y_{\ell m}\left(\theta,\phi\right)$, lose their orthogonality. This
leads to the fourth and last contaminant, which is confusion from
the higher-order CMB temperature fluctuations when analysing
partial-sky observations.

\begin{figure}[t]
	\centering
	\begin{subfigure}[b]{0.24\textwidth}
		\centering
		\includegraphics[width=\textwidth]{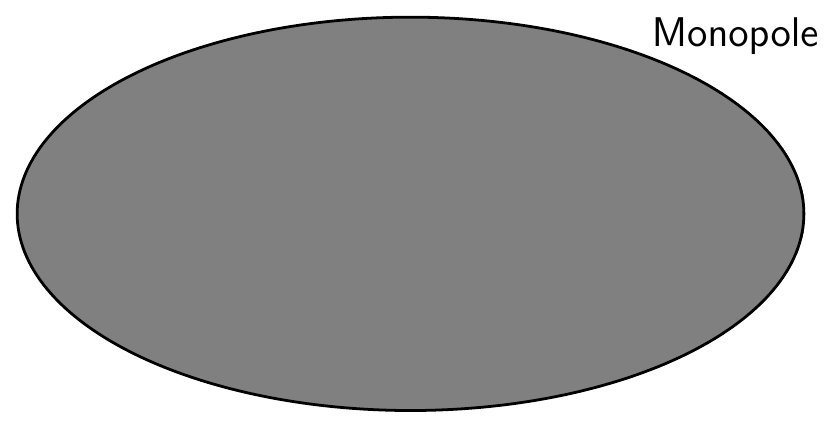}
	\end{subfigure}
	\begin{subfigure}[b]{0.24\textwidth}
		\centering
		\includegraphics[width=\textwidth]{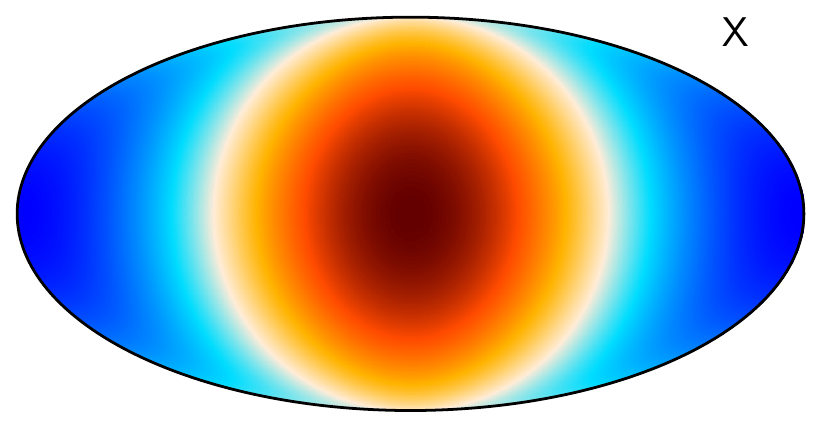}
	\end{subfigure}
	\begin{subfigure}[b]{0.24\textwidth}
		\centering
		\includegraphics[width=\textwidth]{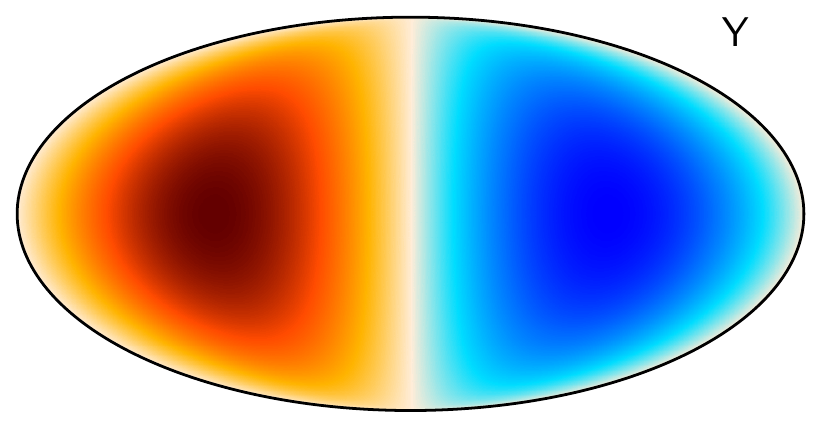}
	\end{subfigure}
	\begin{subfigure}[b]{0.24\textwidth}
		\centering
		\includegraphics[width=\textwidth]{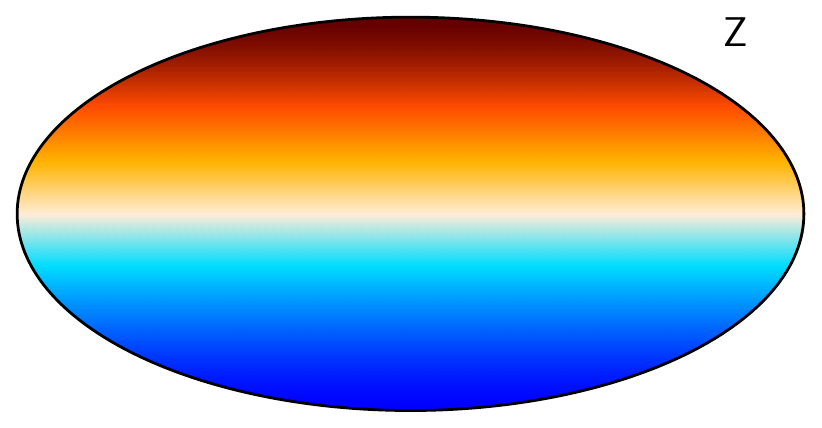}
	\end{subfigure}
	\caption{Basic templates used for CMB dipole estimation in the
          traditional template fitting approach. These correspond to
          the monopole (\emph{upper left}) and the $x$- (\emph{upper right}), $y$-
          (\emph{lower left}), and $z$-dipoles (\emph{lower right}).}
	\label{fig:templates}
\end{figure}

The traditional way of CMB dipole parameter estimation with real
data has typically followed a fairly simple approach. First, the
orbital dipole contribution is removed from the time-ordered data of a
given experiment, which are subsequently co-added into pixelized sky
maps. Then foreground contamination is suppressed either through some
component separation technique or by simple template regression with
respect to known foreground tracers. Next, some part of the sky is
removed by masking, before finally the CMB dipole is estimated
through template fitting with partial-sky and foreground-cleaned data,
typically adopting the templates shown in
Fig.~\ref{fig:templates}. (Note that the monopole is usually included
in the fit for marginalization purposes only, to avoid potential
inaccuracies in the zero-level determination from biasing the dipole
fit.) For one specific example of such an implementation, see, e.g.,
\citet{planck2014-a03}.

As instrumental sensitivity has improved through the years, the
specific details of each step in this procedure have become more
important. For instance, for COBE-FIRAS \citep{fixsen:1994} the
uncertainty on the dipole amplitude due to statistical noise was
6\muK, while the foreground-induced uncertainty was about 14\muK. For
comparison, \citet{lineweaverdipole} estimated that the uncertainty
due to confusion from higher-order CMB fluctuations was 3\muK, and
this particular term was therefore irrelevant for COBE.

The same did not hold true for the Wilkinson Microwave Anisotropy
Probe (WMAP) experiment, for which both the raw sensitivity and
foreground rejection capabilities improved massively with respect
to COBE. For this reason, the WMAP team replaced the simple dipole
template-fitting procedure discussed above with a more sophisticated
and optimal Wiener filter method \citep{hinshaw2009} originally
pioneered by
\citet{jewell:2004,wandelt:2004,PowerSpectrumEstimation}. The main
advantage of this approach is the fact that the higher-order CMB
fluctuations are estimated jointly with the dipole parameters, and by
assuming that these correspond to a statistically isotropic and
Gaussian random field, it is possible to partially reconstruct their
properties even inside the Galactic mask.

The main goal of the current paper is to quantify the relative
performance of the template fitting and Wiener filter methods. This
has recently become a particularly important topic in the context of
the \Planck\ experiment, for which the sensitivity and control of
systematic effects is so high that the total error budget has now
become dominated by the higher-order CMB contribution. Specifically,
the total instrumental uncertainty on the dipole amplitude is about
1\muK\ \citep{planck2016-l01}, whereas the CMB confusion term arising
from the naive template approach is, as we will see, typically
between 1 and 3\muK, depending on sky fraction. Minimizing this term
is therefore critically important. The results we obtain when applying
this analysis framework to the latest \Planck\ observations are
summarized in \citet{npipe}.

The rest of this paper is organized as follows: In
Sec.~\ref{sec:theory} we give a short theoretical introduction on how
the dipole parameters are obtained from sky maps, what effect partial
sky coverage has on the uncertainties of these parameters, and further
present a Wiener filter method for the estimation of these
uncertainties. In Sec.~\ref{sec:implementation} we describe the
implementation of the uncertainty estimation technique. We present our
results in Sec.~\ref{sec:results}, and we give our conclusions in
Sec.~\ref{sec:conclusions}.

\section{Notation and methods}
\label{sec:theory}
We start our discussion with a review of the traditional approach for
estimating the dipole parameters (amplitude $A$, galactic longitude
$l$ and latitude $b$) from a CMB map, and a discussion of how partial
sky coverage complicates this procedure.

In this paper, we take as a starting point for dipole parameter
estimation a cleaned CMB map in which as much foreground emission as
possible has been removed; for details on how to perform foreground
cleaning, we refer the interested reader to, e.g.,
\citet{planck2016-l04} and references therein. However, no component
separation technique allows for perfect foreground removal, and there
will always be some residual emission, especially in the galactic plane. One therefore must typically
apply a mask to eliminate heavily contaminated regions on the
sky. How to optimally estimate the dipole parameters from such a
clean but incomplete CMB map is the main topic of this section. 

\subsection{Traditional template fitting}
\label{sec:template_fitting}

As a starting point for our analysis, we assume that the clean sky map
may be written in the form
\begin{equation}
  \d = \T\a + \n,
\end{equation}
where $\T$ is a matrix containing the monopole and dipole templates in
its columns, shown in Fig.~\ref{fig:templates}; $\a$ is a
corresponding vector of template amplitudes; and $\n$ is noise. Note
that the latter may or may not include higher-order CMB
fluctuations. In addition, we define a noise covariance matrix
$\N=\left<\n\n^t\right>$, and a diagonal mask matrix $\M$ that is zero
for masked pixels and unity for unmasked pixels.

\begin{figure}[t]
	\centering
	\begin{subfigure}[b]{\columnwidth}
		\centering
		\includegraphics[width=\textwidth]{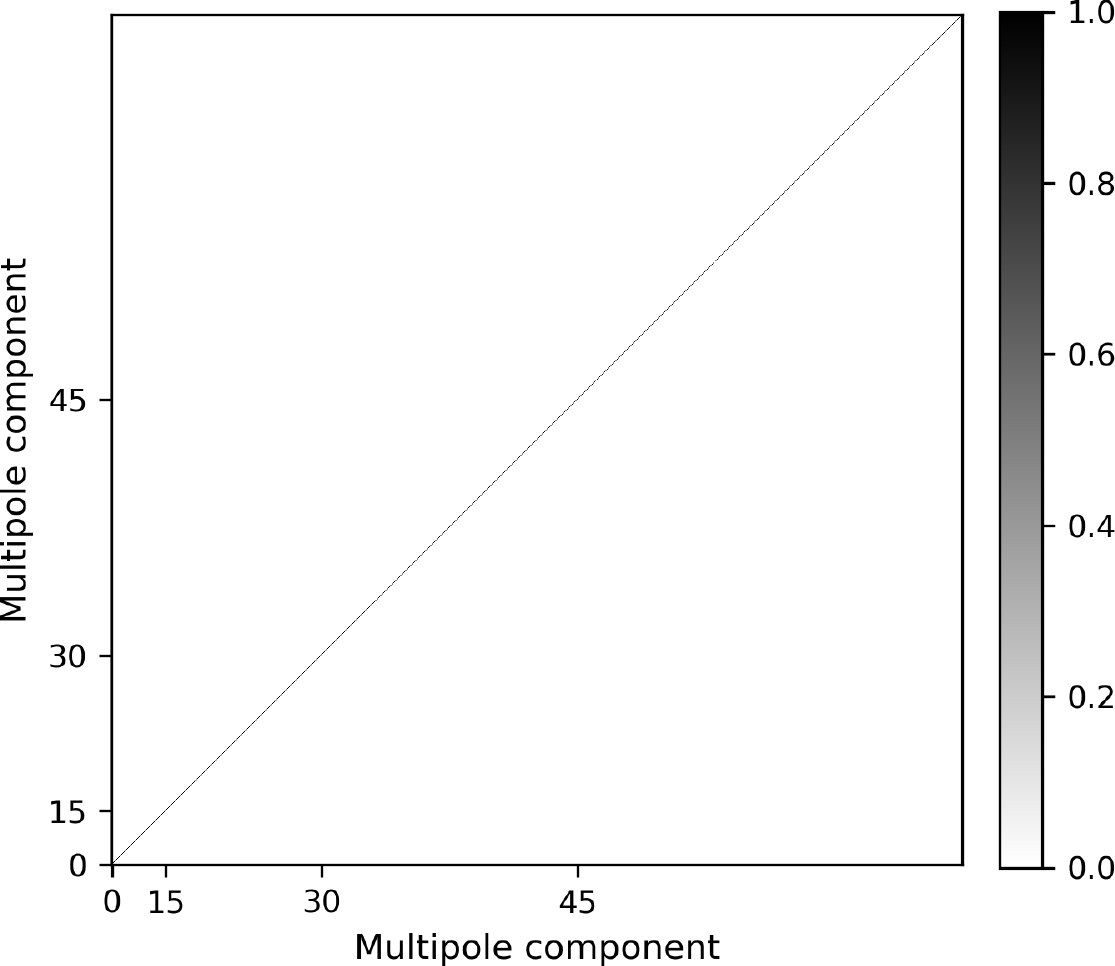}
	\end{subfigure}
	
	\begin{subfigure}[b]{\columnwidth}
		\centering
		\includegraphics[width=\textwidth]{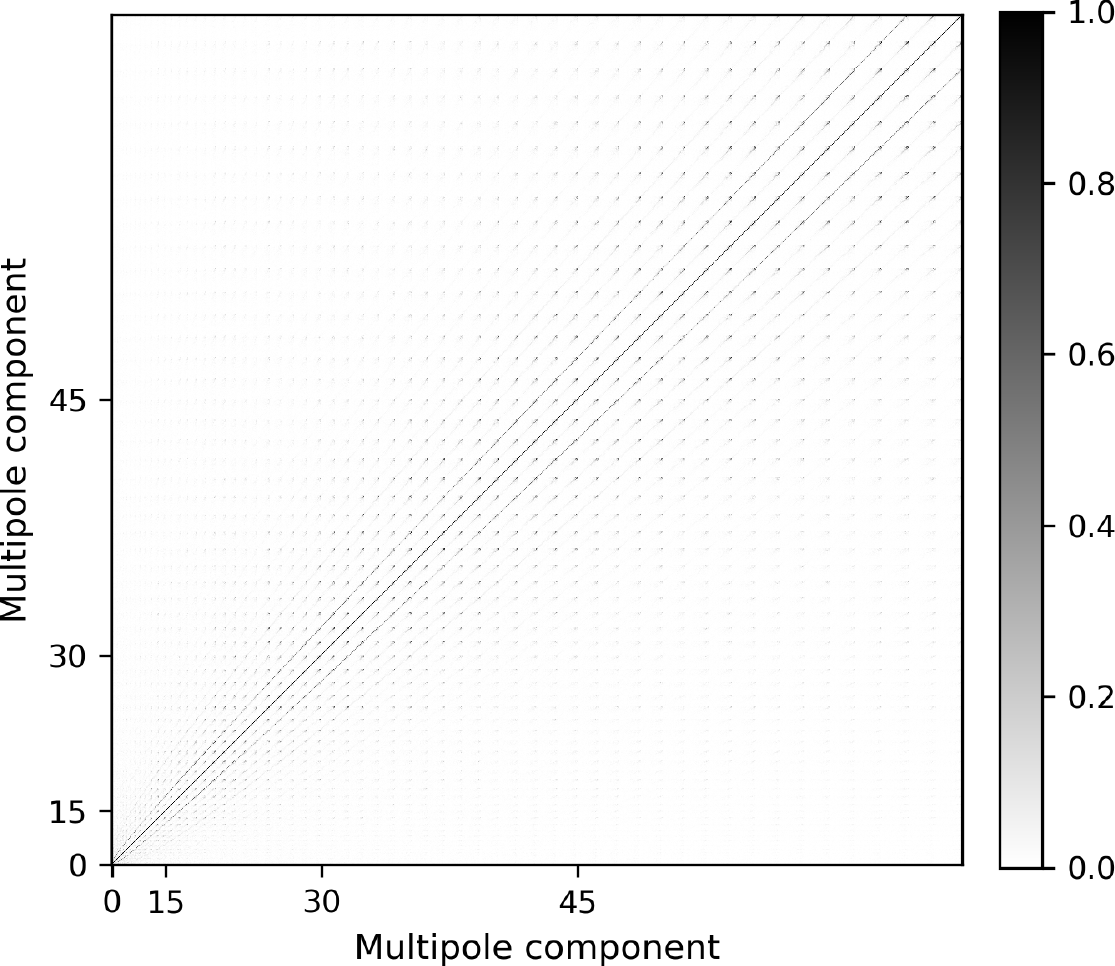}
	\end{subfigure}
	\caption{Spherical harmonics coupling kernel $K$ for $\ell \le 60$
          for an unmasked (\emph{top}) and a masked (\emph{bottom})
          sky. Multipole moments are listed in $\ell$-major ordering
          with element numbering given by $i=\ell^2+\ell+m+1$. The sky
          fraction for the masked case is
          36~\%.}
	\label{fig:coupling_kernel}
\end{figure}

\begin{figure}[t]
	\centering
	\begin{subfigure}[b]{\columnwidth}
		\centering
		\includegraphics[width=\textwidth]{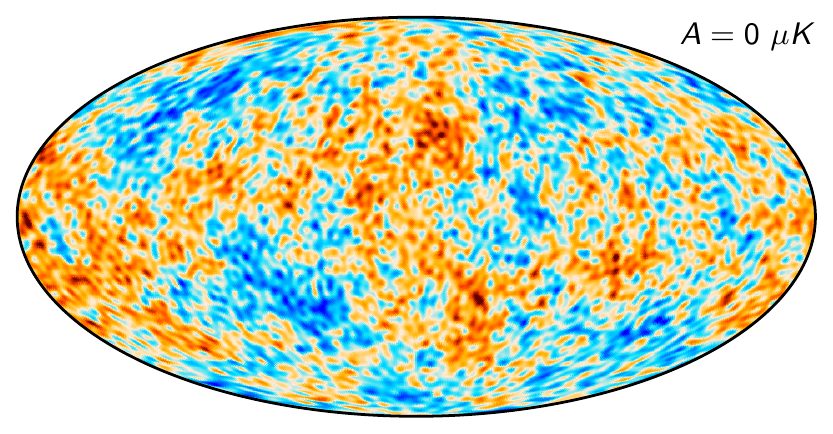}
	\end{subfigure}

	\begin{subfigure}[b]{\columnwidth}
		\centering
		\includegraphics[width=\textwidth]{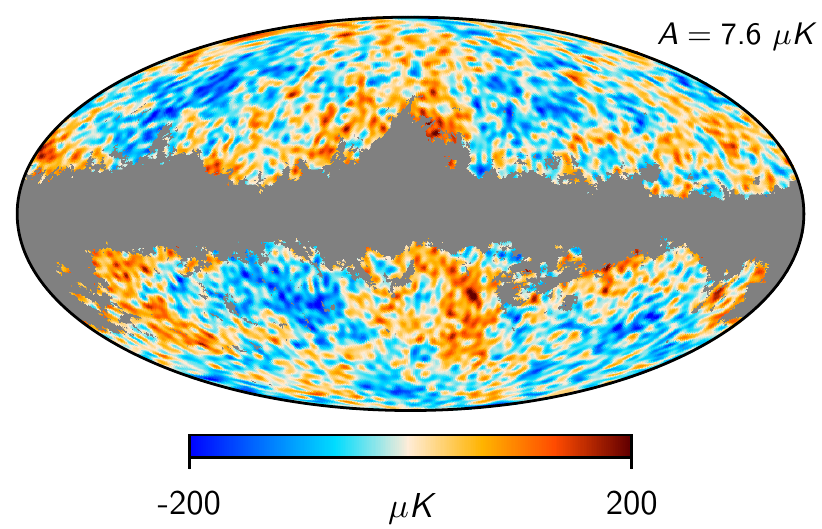}
	\end{subfigure}
	\caption{Illustration of spurious dipole excited from
          higher-order moments through mask coupling. The top panel
          shows an ideal CMB sky with identically vanishing dipole
          moment. The bottom panel shows the same realization, but
          with the Galactic plane masked. By accident, one large
          extended hotspot happens to lie on the western hemisphere
          inside the mask. Once this is removed by the mask, the net
          unmasked result is a dipole pointing toward the Eastern
          hemisphere with an amplitude of 7.6\muK.}
	\label{fig:partialsky}
\end{figure}

Assuming that the noise is Gaussian, the
maximum-likelihood solution for $\a$ is then given by the so-called
``normal equations'',
\begin{equation}
\a = {\left(\T^t\M\N^{-1}\T\right)}^{-1}\T^t\M\N^{-1}\d.
\label{eq:templatefit}
\end{equation}
The trigonometric relations between these amplitudes and the dipole parameters are
\begin{align}
	A&=\sqrt{a_x^2+a_y^2+a_z^2} \label{eq:amp} \\
	l&=\arctan(a_y/a_x) \label{eq:lon} \\
	b&=90-\arccos(a_z/A). \label{eq:lat}
\end{align}
Here, $a_x$, $a_y$ and $a_z$ are the components of the coefficient vector $\a$ and $A$, $l$ and $b$ are the dipole amplitude, longitude and latitude respectively. Note that many commonly used implementations of this approach do not
implement full inverse variance noise weighting, as described by
Eq.~\ref{eq:templatefit}, but simply adopt $\N=\mathds{1}$, and thereby in
effect assign equal weight to all pixels. We will do the same in the
following.

\subsection{Complications from partial sky coverage}
\label{sec:partial_sky_coverage}
The dipole parameters in Eq.~\ref{eq:amp}--\ref{eq:lat} are subject to
confusion from small-scale CMB fluctuations whenever a mask is
applied. To see this, we expand the CMB fluctuation field into
spherical harmonics as follows,
\begin{equation}
T\left(\hat{n}\right) =
\sum_{\ell=0}^{\ell_{max}}\sum_{m=-\ell}^{\ell}a_{\ell m}Y_{\ell m}\left(\hat{n}\right),
\label{eq:SphHarExpansion}
\end{equation}
where $T$ is the CMB fluctuation map; $Y_{\ell m}$ are the spherical
harmonic base functions; and the $a_{\ell m}$ are the associated
weights. With access to the full celestial sphere, these coefficients
may be computed as
\begin{equation}
a_{\ell m}=\int_{4\pi}T\left(\hat{n}\right)Y_{\ell m}^*\left(\hat{n}\right)d\Omega.
\end{equation}
However, when masking parts of the sky with a mask $\M$ the spherical harmonic base
functions are no longer orthogonal, and the new so-called
pseudo-harmonic coefficients read \citep{hivon:2002}
\begin{equation}
\begin{split}
\tilde{a}_{\ell m} & =\int_{4\pi}\M T\left(\hat{n}\right)Y_{\ell m}^*\left(\hat{n}\right)d\Omega \\
& = \sum_{\ell^{'}m^{'}} a_{\ell^{'}m^{'}}K_{\ell m,\ell^{'}m^{'}}\left[\M\right].
\end{split}
\end{equation}
Here, $K$ is called the coupling kernel, and quantifies the
mutual dependence between any two modes $Y_{\ell m}$ and $Y_{\ell'm'}$. The
explicit expression for the coupling kernel reads
\begin{equation}
K_{\ell m,\ell^{'}m^{'}}= \int_{4\pi}^{}\M\left(\hat{n}\right)Y_{\ell m}\left(\hat{n}\right)Y_{\ell^{'}m^{'}}^*\left(\hat{n}\right)d\Omega.
\label{eq:coupling_kernel}
\end{equation}
Figure~\ref{fig:coupling_kernel} shows $K$ for multipoles up to
$\ell_{\text{max}}=60$ for two different cases. In the no-mask case,
shown in the top panel, all modes are orthogonal and therefore
independent. When we apply a mask, shown in the bottom panel, non-zero
off-diagonal elements appear. In other words, any higher-order mode
will induce a spurious dipole contribution unless properly accounted
for.

Figure~\ref{fig:partialsky} provides an intuitive illustration of
this effect. The top panel shows an ideal CMB realization drawn from a
$\Lambda$CDM power spectrum with an identically vanishing dipole
moment, $A=0$. However, one of the largest hotspots on the sky
happens, by chance, to align closely with the western half of
the Galactic plane. After applying the Galactic mask, which gives this
hotspot zero weight in the dipole fit, the result is a net dipole that
points in the opposite direction. In this particular case, the result
is a spurious dipole of $7.6\muK$ pointing towards the eastern
hemisphere.

\subsection{Wiener filtering and Gibbs sampling}
\label{sec:Wiener_filtering_approach}

\begin{figure}[t]
	\centering
	\begin{subfigure}[b]{\columnwidth}
		\centering
		\includegraphics[width=\textwidth]{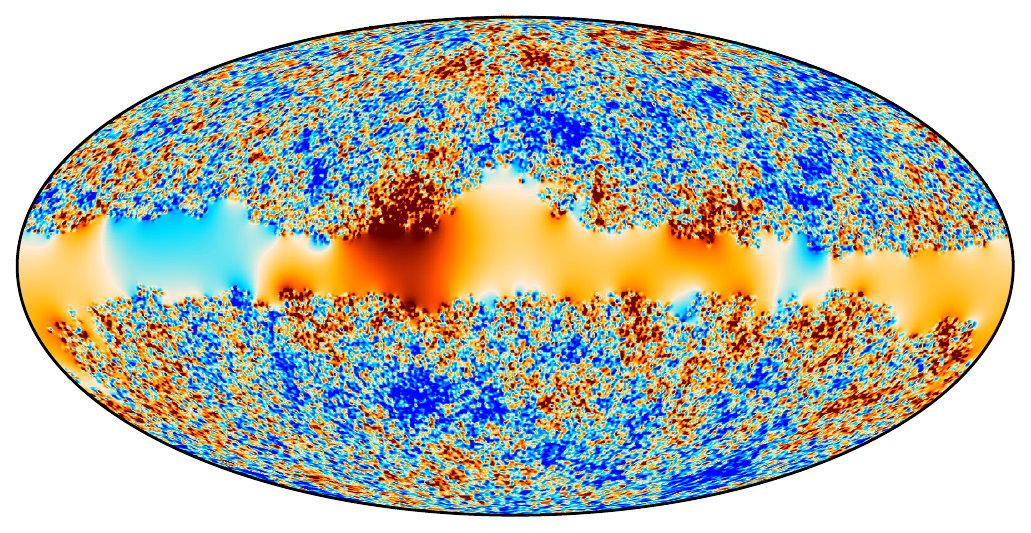}
	\end{subfigure}
	
	\begin{subfigure}[b]{\columnwidth}
		\centering
		\includegraphics[width=\textwidth]{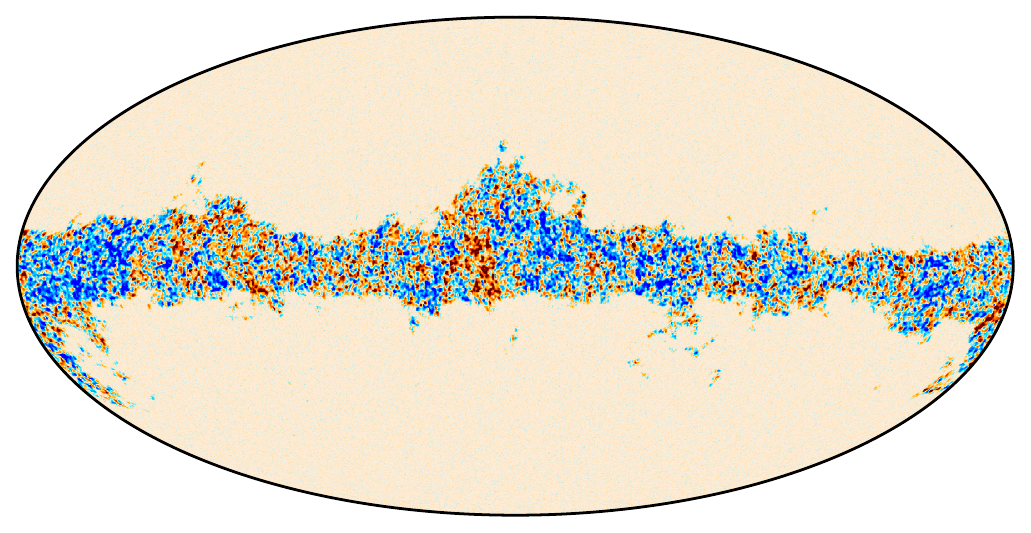}
	\end{subfigure}

	\begin{subfigure}[b]{\columnwidth}
		\centering
		\includegraphics[width=\textwidth]{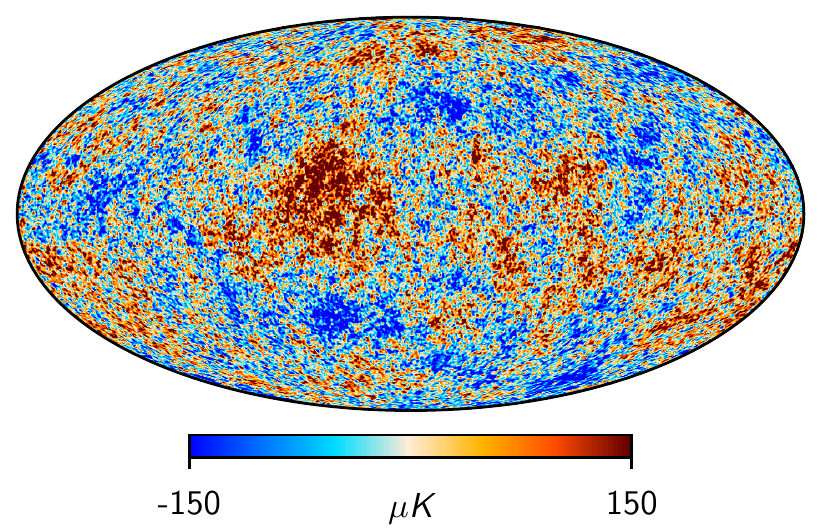}
	\end{subfigure}
	\caption{Illustration of the Gibbs sampling procedure. The
          full sample $\s$ (\emph{bottom panel}) is the sum of the mean
          field map ${\hat{\x}}$ (\emph{top panel}) and a
          fluctuation map ${\hat{\y}}$ (\emph{middle panel}).}
	\label{fig:Wienersamples}
\end{figure}

Within the simple template fitting approach described above, the
higher-order CMB fluctuations are treated as a noise term. Since the
CMB fluctuations are Gaussian and isotropic, this term does not lead
to any bias in the central estimates, but it does increase the
variance. An alternative approach is to exploit the assumptions of
isotropy and Gaussianity to estimate the CMB signal jointly with the
dipole parameters, adopting the following data model,
\begin{equation}
  \d = \s + \n,
  \label{eq:basic_model1}
\end{equation}
where
\begin{equation}
  \s = \s(\hat{n}) =\sum_{\ell,m} a_{\ell m} Y_{\ell m}(\hat{n})
  \label{eq:basic_model2}
\end{equation}
now is an isotropic and Gaussian random field with some angular
power spectrum $C_{\ell}$. Note that the previously defined dipole is
contained within $\s$ in the form of $\T\a = \sum_{\ell = 0}^1
\sum_{m} a_{\ell m} Y_{\ell m}(\hat{n})$. 

To estimate $\s$ and $C_{\ell}$ jointly, we employ the Gibbs sampling
algorithm described by \cite{PowerSpectrumEstimation}. This algorithm
draws samples from the probability density $P(\s,C_\ell|\d)$. Since it
is difficult to sample directly from this joint distribution, we
instead employ Gibbs sampling, and perform consecutive sampling from
each conditional density, $P(\s|C_\ell,\d)$ and $P(C_\ell | \s,\d)$,
which according to Gibbs sampling theory will converge to being
samples from the joint density $P(\s,C_\ell|\d)$. Thus, the two Gibbs
sampling steps are
\begin{align}
\s^{i+1}&\leftarrow P(\s|C_\ell^i,\d), \label{eq:sample_map}\\
C_\ell^{i+1}&\leftarrow P(C_\ell|\s^{i+1}). \label{eq:Cl_sampling}
\end{align}
For more information on the sampling process for the CMB power
spectrum in Eq.~\ref{eq:Cl_sampling}, we again refer to
\cite{PowerSpectrumEstimation}. However, we note that the
inverse-gamma sampler described in that paper is only employed for
multipoles $\ell\ge2$ in the current analysis. For the first two
elements, we manually set $C_\ell$ to a numerical large value of
$10^{12}\muK^2$, which effectively corresponds to imposing no
informative priors on the monopole and dipole moments.

In our context we are mostly interested in the map sampling process in
Eq.~\ref{eq:sample_map}. In effect, the sky sample $\s^i$ uses phase
information in the data $\d$ outside the mask to extrapolate into the
missing pixels. The result is a constrained realization with the
assumed power spectrum $C_\ell$, such that the full map is a sample
from the desired target distribution.

The map sampling process in Eq.~\ref{eq:sample_map} is performed in
two steps. First we compute the so-called \textit{mean field map} by
solving the Wiener filter mapmaking equation for $\hat{\x}$,
\begin{equation}
\left(\S^{-1}+\Y^t\N^{-1}\Y\right){\hat{\x}}=\Y^t\N^{-1}\d.
\end{equation}
Here, $\S$ is a diagonal prior matrix that contains the assumed power
spectrum, $\Y$ denotes spherical harmonic transforms, and $\N$ is the
noise covariance matrix. The equation above is a Wiener filter, and the
resulting map ${\hat{\x}}$ is as such biased. To unbias the
full sample, we have to add a fluctuation term. We obtain the
corresponding \textit{fluctuation map} by solving the following
expression for ${\hat{\y}}$,
\begin{equation}
\left(\S^{-1}+\Y^t\N^{-1}\Y\right){\hat{\y}}=\S^{-1/2}\mathbf{\omega_1}+\Y^t\N^{-1/2}\mathbf{\omega_2}.
\end{equation}
Here, $\mathbf{\omega_1}$ and $\mathbf{\omega_2}$ are two independent Gaussian white
noise maps with zero mean and unit variance. The full sample
$\s$ is then the sum of ${\hat{\x}}$ and
${\hat{\y}}$.

\begin{figure}[t]
	\centering
	\includegraphics[width=\columnwidth]{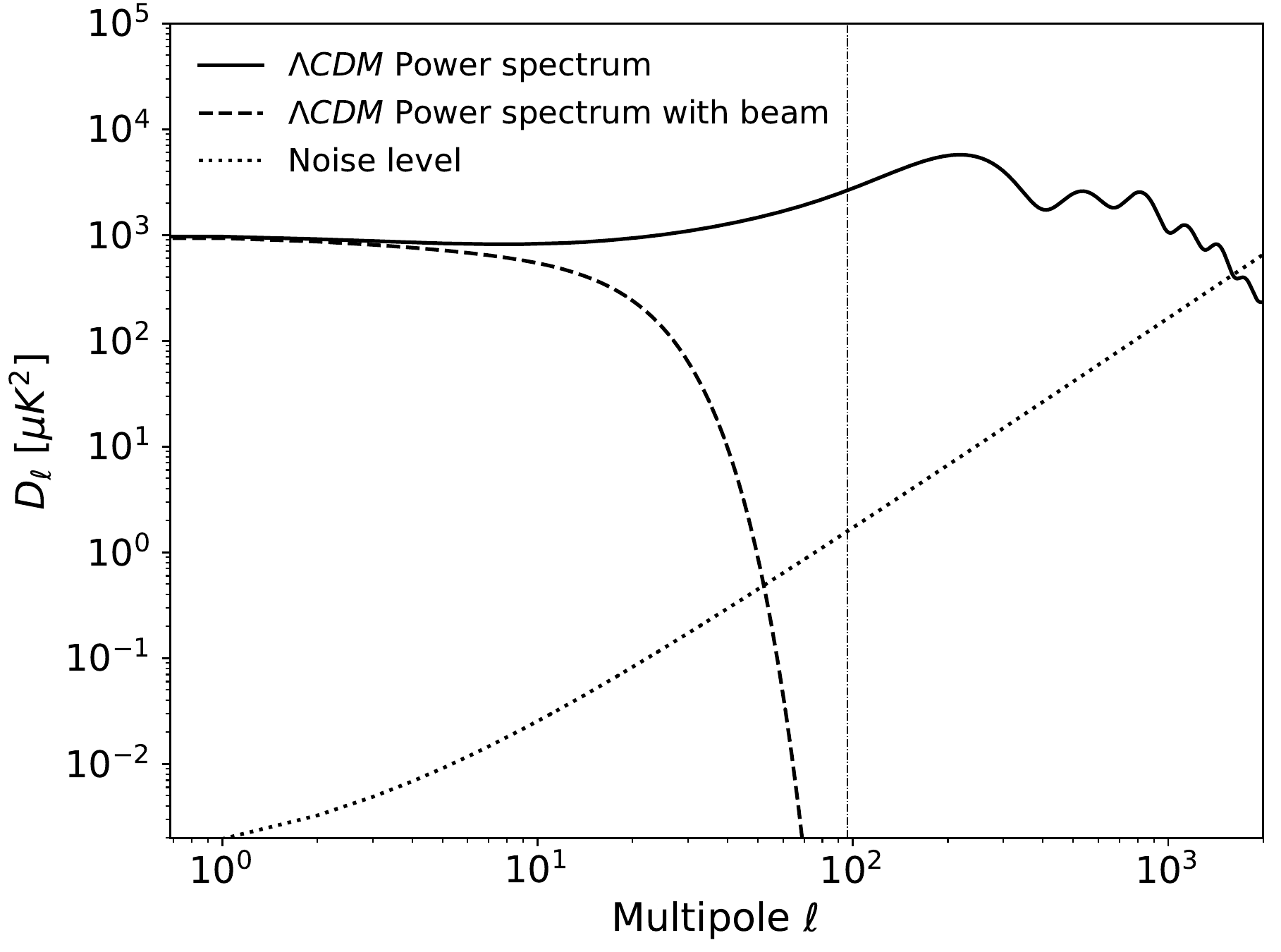}
	\caption{Overview of signal and noise power spectra adopted for
          the Monte Carlo simulations. The solid line shows the
          best-fit \Planck\ $\Lambda$CDM power spectrum, and the
          dashed line shows the same, but convolved with a $6^{\circ}$
          FWHM Gaussian beam. The dotted line shows the noise spectrum
          for white uncorrelated noise with $\sigma_0=1\muK$ per pixel
          at a HEALPix resolution of ${\nside}=32$. The vertical
          dotted line shows the harmonic space truncation limit of
          $\ell_{\textrm{max}}=95$.}
	\label{fig:noise_level}
\end{figure}

The key advantage of sampling from the full distribution is that one
simultaneously takes into account all multipole scales and all elements
of the coupling kernel shown in Fig.~\ref{fig:coupling_kernel}. To
illustrate this visually, Fig.~\ref{fig:Wienersamples} shows the
different maps involved in the sampling procedure for some typical
mask. The top panel shows the Wiener filter component. Note that this
map contains small-scale structures in the unmasked regions,
but only smooth structures in the masked regions. However, critically,
it is not zero inside the mask. On the contrary, because of the
assumptions of statistical isotropy and Gaussianity, the field inside
the mask must show some degree of phase correlation with the unmasked
regions, and this is precisely the information that allows partial
reconstruction inside the mask.

Of course, this extrapolation is only supported for large angular
scales. The fluctuation term, shown in the middle panel, therefore
compensates for the fluctuation power that is lost due to the mask and
noise, such that the sum of the two components, shown in the bottom
panel, is a single full-sky map that is consistent with the original
data. Note, however, that this map contains a significant stochastic
component, and a full ensemble of such Gibbs samples is therefore
required to adequately describe both the mean and covariance of the
true underlying signal.

\section{Simulations}
\label{sec:implementation}

We have now established two different methods for estimating dipole
parameters from a foreground-cleaned CMB map with partial sky
coverage. In order to assess the relative performance of these two
methods, we perform Monte Carlo simulations for both algorithms, and
compare the resulting uncertainties.

\subsection{Monte Carlo procedure}

For the standard template fitting approach, the procedure is defined
as follows:
\begin{enumerate}
\item Generate $N$ simulated CMB skies $d_{[1,...,N]}$ based on
  a $\Lambda$CDM power spectrum.
\item Apply mask $\M$ to each sample.
\item Compute best-fit dipole parameters $A$, $l$ and $b$ for each sample with
  Eqs.~\ref{eq:templatefit}--\ref{eq:lat}.
\item Report the standard deviation for each parameter, $\sigma_A$, $\sigma_l$ and $\sigma_b$. 
\end{enumerate}

For the Wiener filter approach, the procedure is similar, but
additionally involves an intermediate sampling loop for each realization:
\begin{enumerate}
\item Generate $N$ simulated CMB skies $d_{[1,...,N]}$ based on
  a $\Lambda$CDM power spectrum.
\item Apply mask $\M$ to each sample.
\item For each masked CMB map: 
\item[]{\begin{enumerate}
  \item Draw $n$ full-sky Wiener filter samples $s_{[1,...,n]}$.
  \item Compute dipole parameters $A$, $l$ and $b$ for each sample.
  \item Compute single-realization standard deviations $\sigma_A^i$, $\sigma_l^i$ and $\sigma_b^i$.
\end{enumerate}}
\item Report the mean of $\sigma_A^i$, $\sigma_l^i$ and $\sigma_b^i$.
\end{enumerate}

The template fitting analysis is implemented using the HEALPix
\texttt{fit\_dipole} routine \citep{gorski:2005}, while the Wiener
filter analysis is performed using the \texttt{Commander} code
\citet{PowerSpectrumEstimation}.

\begin{figure}[ht!]
	\centering
	\begin{subfigure}[b]{0.90\columnwidth}
		\centering
		\includegraphics[width=\textwidth]{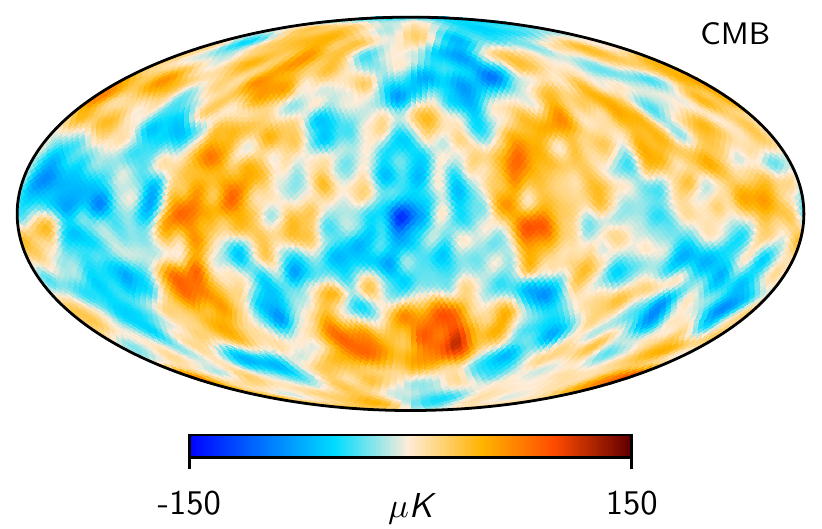}
	\end{subfigure}
	
	\begin{subfigure}[b]{0.90\columnwidth}
		\centering
		\includegraphics[width=\textwidth]{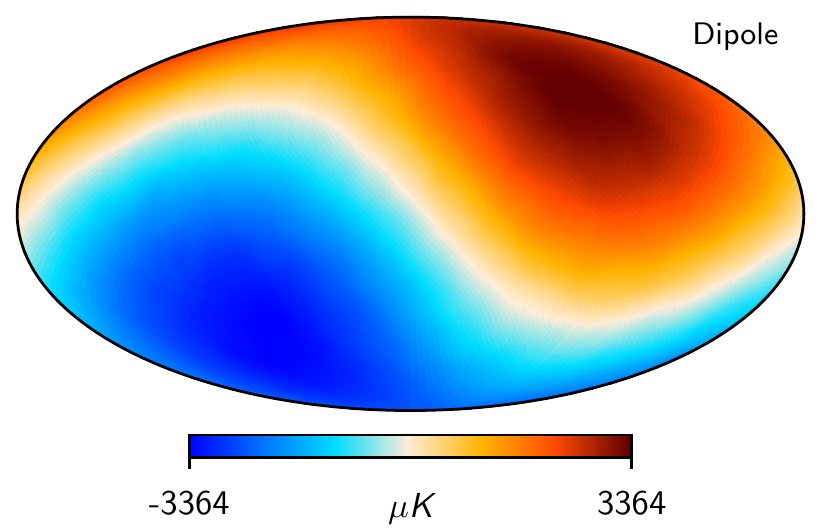}
	\end{subfigure}
	
	\begin{subfigure}[b]{0.90\columnwidth}
		\centering
		\includegraphics[width=\textwidth]{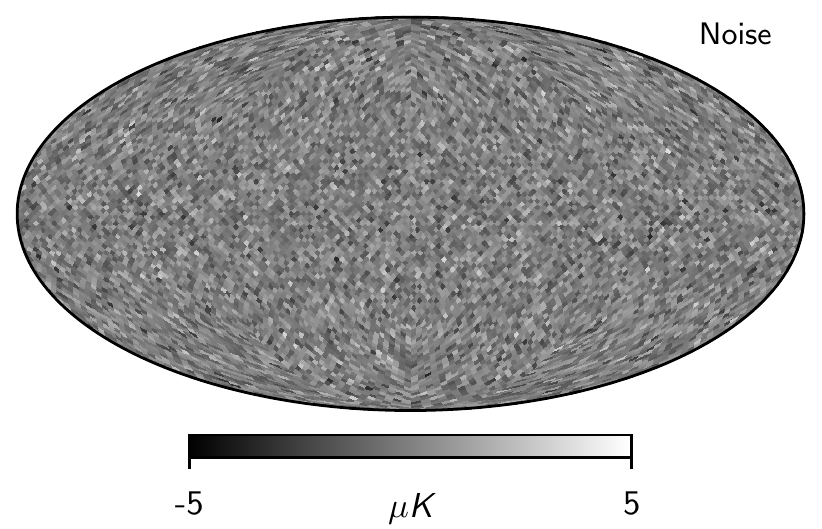}
	\end{subfigure}
	
	\begin{subfigure}[b]{0.90\columnwidth}
		\centering
		\includegraphics[width=\textwidth]{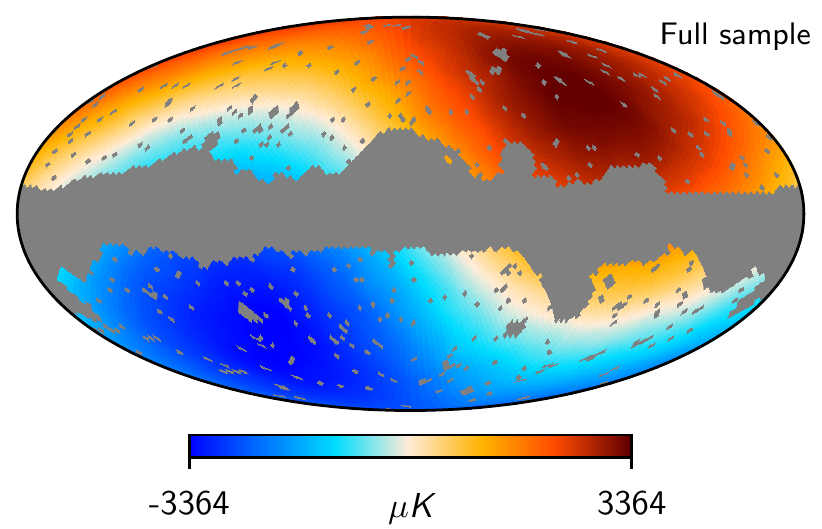}
	\end{subfigure}
	\caption{Components used to construct each Monte Carlo
          realization. These are the higher-order CMB fluctuations (\emph{top
            panel}), the CMB dipole (\emph{second panel}), and
          instrumental noise (\emph{third panel}). The bottom panel
          shows the sum of the three components with a mask
          super-imposed.}
	\label{fig:CMBsimulation}
\end{figure}

\begin{figure}[t]
	\centering
	\includegraphics[width=0.9\columnwidth]{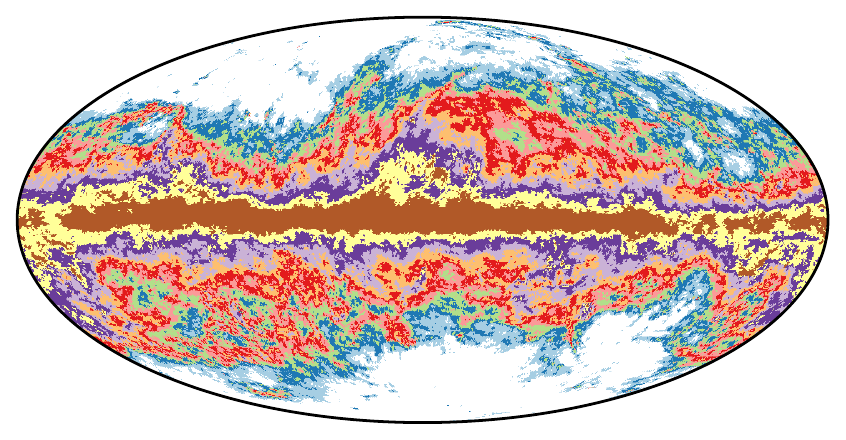}
	\caption{Analysis masks used for the Monte Carlo comparison. The unmasked sky fractions from the smallest to the largest mask are 92\,\% (brown), 84\,\% (yellow), 76\,\% (purple), 68\,\% (lilac), 60\,\% (bright orange), 52\,\% (red), 44\,\% (bright red), 36\,\% (green), 28\,\% (cyan) and 20\,\% (light blue).}
	\label{fig:joint_mask}
\end{figure}

\subsection{CMB simulations}
\label{sec:simulations}

For our simulated sky maps, we adopt the same model as in
Eqs.~\ref{eq:basic_model1}--\ref{eq:basic_model2}, but this time we
explicitly include support for an instrumental beam with Legendre
expansion $b_{\ell}$,
\begin{equation}
\d = \s+\n = \sum_{\ell=0}^{\ell_{max}}\sum_{m={-\ell}}^{\ell}b_{\ell} a_{\ell m} Y_{\ell m} + \n.
\label{eq:datamodel}
\end{equation}
We generate multipoles with $\ell\ge2$ using Healpy's
\texttt{synfast}\footnote{http://healpix.jpl.nasa.gov} routine, drawn
from the \Planck\ 2018 best-fit $\Lambda$CDM power spectrum
\citep{planck2016-l06}. For the dipole component, we adopt
$A=3364.0\muK$, $l=264.1\deg$ and $b=48.3\deg$. No monopole is added.

For both the template and Wiener filter approaches, we establish
ensembles of 100 realizations. To limit the computational speed
involved in the Wiener filter stage, which requires iterative
sampling, we choose to perform the analysis at a HEALPix resolution of
$\nside=32$, corresponding to a pixel size of about $1.8\deg$. This is
sufficient to capture all features that are relevant for dipole
estimation.

For high-sensitivity experiments such as \Planck, the direct template
fitting approach is largely insensitive to specific details of the
instrumental noise, as the dominant noise contributor is the CMB
fluctuations, and $\n$ may be safely disregarded in this
framework. However, for the Wiener filter approach some care is
warranted also for this term. In particular, the details of $\n$
determine how aggressively the estimator is able to extrapolate into
the masked region. For the method to be accurate and unbiased, it is
important that the data model in Eq.~\ref{eq:datamodel} actually
is a good representation of the observations in question.

First, since we perform our analysis at $\nside=32$, the highest
resolvable multipole moment is given roughly by
$\ell_{\textrm{max}}\approx3 \nside = 96$. In order to suppress
the signal above this $\ell_{\textrm{max}}$, which is not supported by
Eq.~\ref{eq:datamodel}, we smooth the simulated CMB realizations with
a $6\deg$ FWHM Gaussian beam with a Legendre expansion given by \citep{TegmarkNoise}
\begin{equation}
b_{\ell}=\exp \left[-\frac{1}{2}\ell(\ell+1)\left(\frac{\text{FWHM}\cdot\pi}{180}\cdot \frac{1}{\sqrt{8\ln 2}}\right)^2\right].
\label{eq:beam}
\end{equation}
Next, to avoid ringing artefacts from the truncation limit around the
mask edge, we have to ensure that the effective signal-to-noise ratio is
negligible at $\ell_{\textrm{max}}$. We do this by adding
regularization noise with a standard deviation of $\sigma_0=1\,\muK$
per pixel. In harmonic space, this corresponds to a flat noise
spectrum with an amplitude given by \citep{TegmarkNoise}
\begin{equation}
N_\ell=\sigma_0^2 \cdot \frac{4\pi}{N_{\text{pix}}} \cdot \frac{\ell(\ell+1)}{2\pi}.
\label{eq:noise}
\end{equation}
Figure~\ref{fig:noise_level} summarizes the simulated data in terms of
signal and noise power spectra. Note that with these choices of
parameters, the effective signal-to-noise ratio at $\ell=95$ is
smaller than 0.01. The individual components involved in the simulated map
are illustrated in Fig.~\ref{fig:CMBsimulation}.

Since the elements of the coupling kernel depend on the sky fraction
and on the shape of the mask, we repeat the Monte Carlo analysis for a
variety of different masks with sky fractions ranging from 20 to
95\,\%, shown in Fig.~\ref{fig:joint_mask}. These masks were already
used for a similar purpose in \citet{planck2016-l03}.

\begin{figure}[ht!]
	\begin{centering}
		\includegraphics[width=\columnwidth]{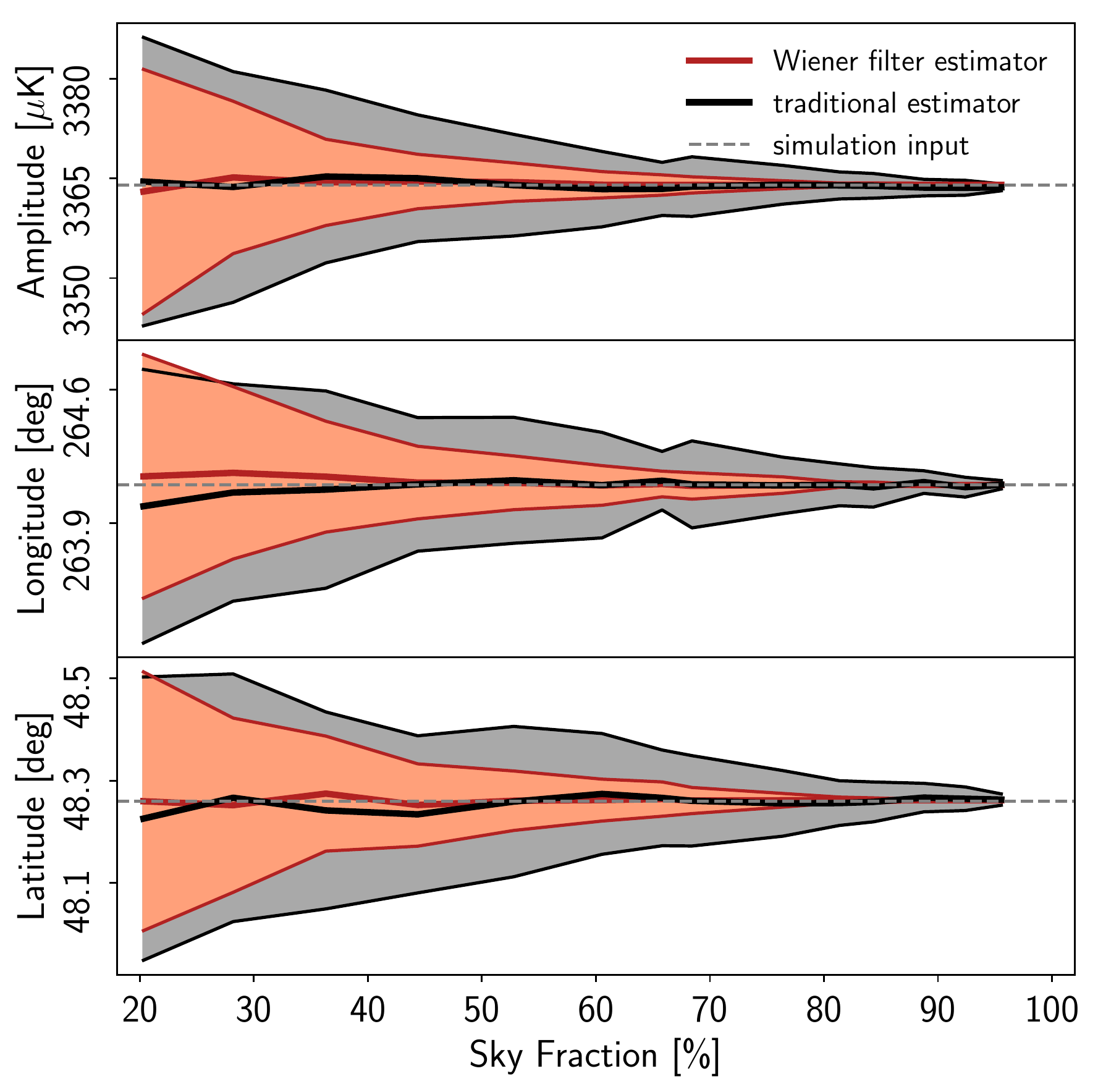}
		\caption{CMB solar dipole parameter uncertainties as a function of sky fraction; The thick red and black lines show the posterior means derived by the Wiener filter method and the traditional method respectively; The shaded bands are the corresponding $\pm1\sigma$ confidence intervals; The horizontal dashed lines mark the true dipole parameters that were used as input for the simulations}
		\label{fig:result}
	\end{centering}
\end{figure}
\begin{figure}[ht!]
	\begin{centering}
		\includegraphics[width=\columnwidth]{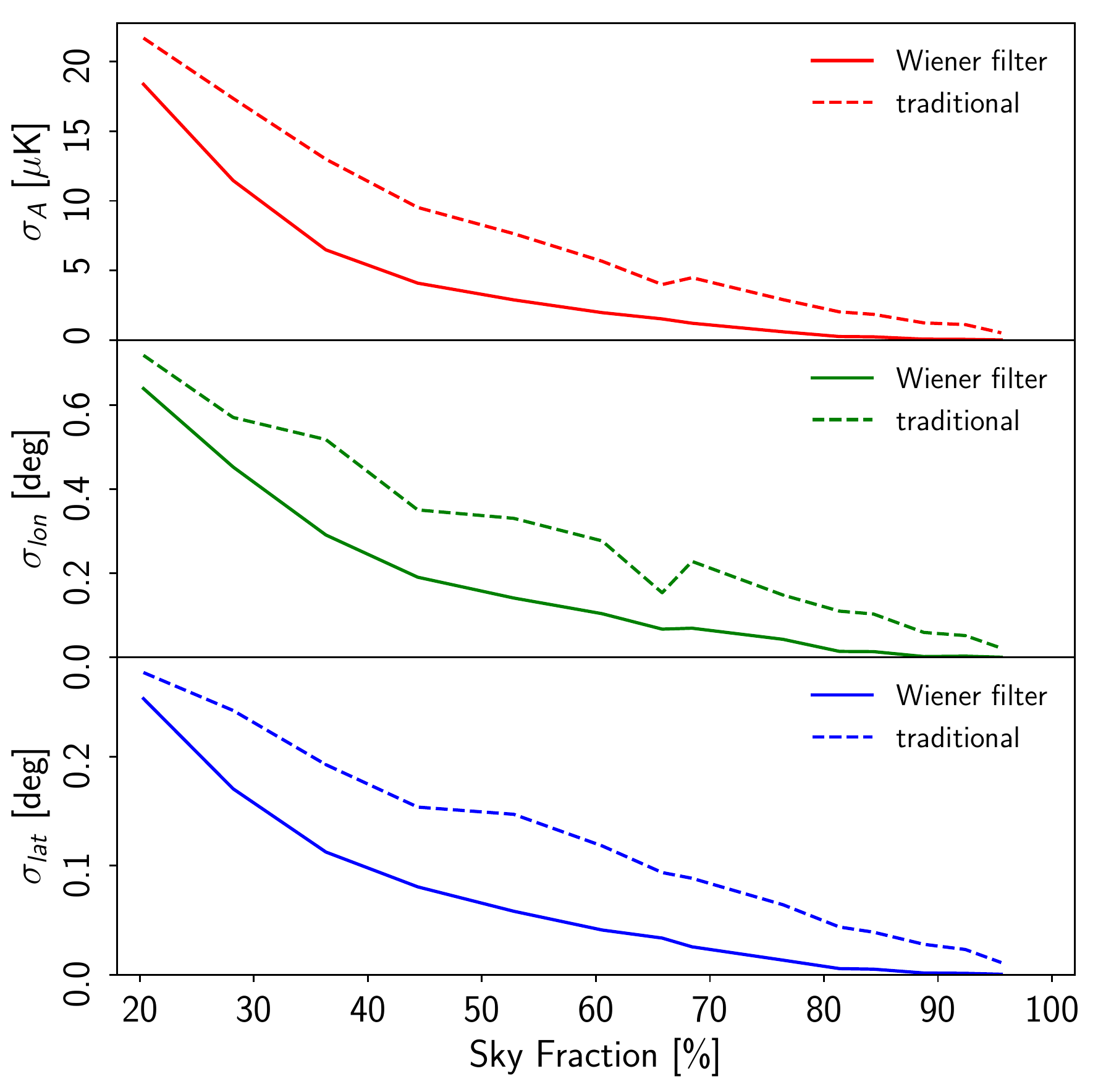}
		\caption{Absolute uncertainties of solar dipole parameters estimated with Wiener filter method (solid lines) and traditional method (dashed lines) as a function of sky fraction}
		\label{fig:sigma_plot}
	\end{centering}
\end{figure}
\begin{figure}[ht!]
	\begin{centering}
		\includegraphics[width=\columnwidth]{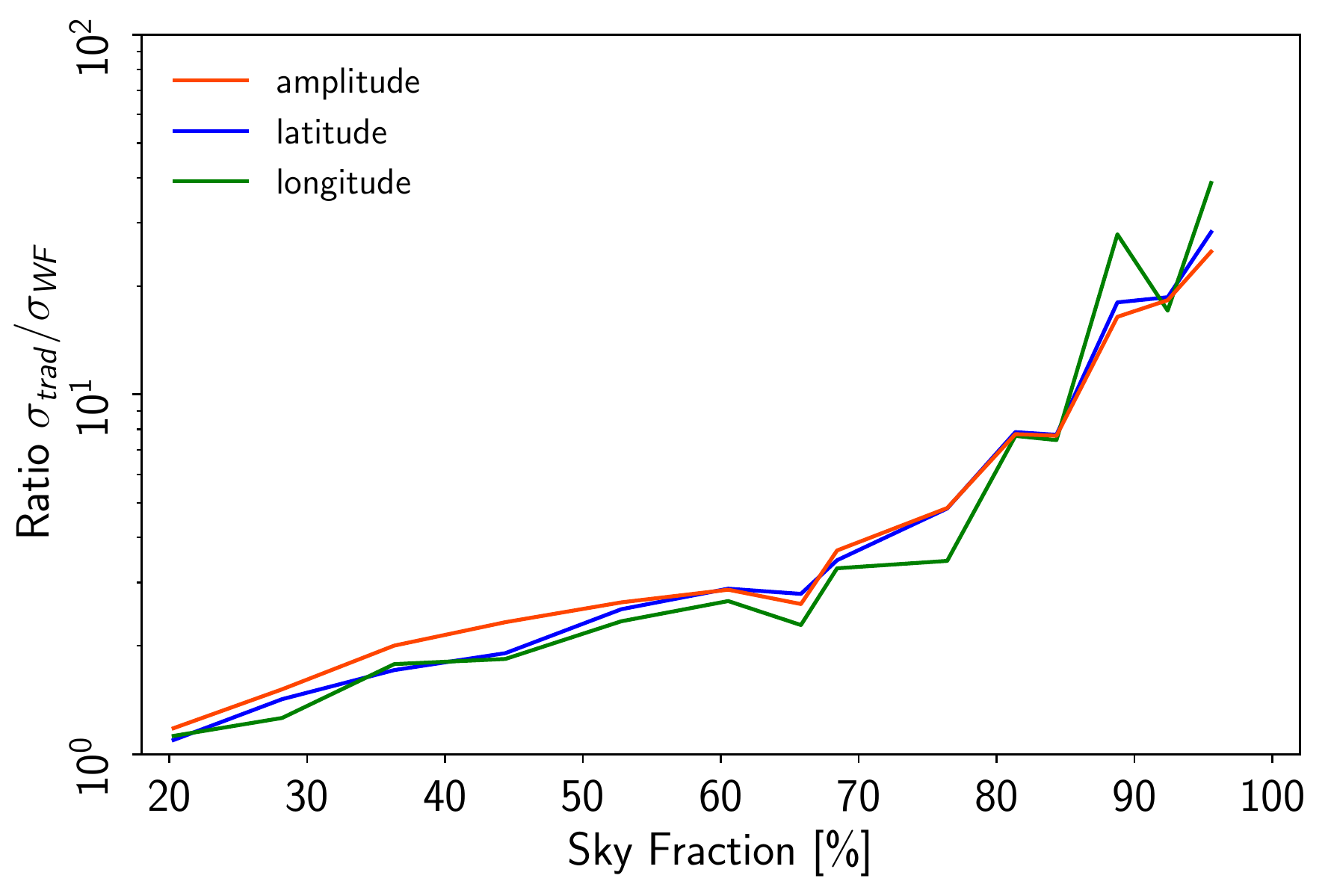}
		\caption{Ratio of uncertainties derived by traditional method and Wiener filter method as a function of sky fraction.}
		\label{fig:ratio}
	\end{centering}
\end{figure}

\section{Results}
\label{sec:results}

We are now ready to present the main result of this paper, which is a
quantitative comparison of the template fitting and Wiener filter
approaches to dipole parameter estimation. For each mask shown in Fig.~\ref{fig:joint_mask}, we analyse 100 independent Monte Carlo
realizations with both methods.

The main result is summarized in Fig.~\ref{fig:result}, where we show
the mean dipole parameters and their corresponding statistical
uncertainties as a function of the sky fraction. From top to bottom,
the panels show the dipole amplitude, the longitude, and the latitude.
The thick red lines show the mean of the derived solar dipole
parameters using the Wiener filter technique, and the regions shaded
in red show the corresponding $\pm1\sigma$ confidence intervals. The
equivalent results derived using the traditional method are shown in
black and gray. We mark the true dipole parameters as horizontal
dashed lines.

We find that the uncertainties derived by the Wiener filter technique
are significantly reduced compared to the traditional method. This
effect is strongest for large sky fractions, for which it is easier to
extrapolate into the masked regions. In contrast, for small sky
fractions the extrapolation is very unreliable, and the two methods
therefore give very similar results.

In some cases in Fig.~\ref{fig:result} it may appear as if the traditional method yields smaller uncertainties than
the Wiener filter at small sky fractions, especially for the longitude
and latitude where the upper confidence interval boundary is in fact
below that of the Wiener filter method. However, this is not actually
the case, since the grey band is also shifted to lower values due to
statistical fluctuations of the derived means. To make this more
explicit, we plot the absolute uncertainties of the two methods in
Fig.~\ref{fig:sigma_plot}, and their ratios in
Fig.~\ref{fig:ratio}. We see that for large sky fractions (above
$\approx 85 \%$) the Wiener filter uncertainties are reduced by a
factor of 10 or more for all parameters. At a sky fraction of about
$50\%$, the uncertainties drop to roughly half of those of the
traditional method.

\section{Conclusions}
\label{sec:conclusions}

\begin{table*}[t]
\newdimen\tblskip \tblskip=5pt
\caption{Comparison of Solar dipole measurements from COBE, WMAP, and \Planck. }
\label{tab:dipole}
\vskip -4mm
\footnotesize
\setbox\tablebox=\vbox{
 \newdimen\digitwidth
 \setbox0=\hbox{\rm 0}
 \digitwidth=\wd0
 \catcode`*=\active
 \def*{\kern\digitwidth}
  \newdimen\dpwidth
  \setbox0=\hbox{.}
  \dpwidth=\wd0
  \catcode`!=\active
  \def!{\kern\dpwidth}
  \halign{\hbox to 2.5cm{#\leaderfil}\tabskip 2em&
    \hfil$#$\hfil \tabskip 2em&
    \hfil$#$\hfil \tabskip 2em&
    \hfil$#$\hfil \tabskip 2em&
    #\hfil \tabskip 0em\cr
\noalign{\doubleline}
\omit&&\multispan2\hfil\sc Galactic coordinates\hfil\cr
\noalign{\vskip -3pt}
\omit&\omit&\multispan2\hrulefill\cr
\noalign{\vskip 3pt} 
\omit&\omit\hfil\sc Amplitude\hfil&l&b\cr
\omit\hfil\sc Experiment\hfil&[\muK_{\rm
CMB}]&\omit\hfil[deg]\hfil&\omit\hfil[deg]\hfil&\hfil\sc Reference\hfil\cr
\noalign{\vskip 3pt\hrule\vskip 5pt}
COBE \rlap{$^{\rm a,b}$}&                  3358!**\pm23!**&     264.31*\pm0.16*&
     48.05*\pm0.09*&\citet{lineweaverdipole}\cr
WMAP \rlap{$^{\rm c}$}&                  3355!**\pm*8!**&     263.99*\pm0.14*&
     48.26*\pm0.03*&\citet{hinshaw2009}\cr
\noalign{\vskip 3pt}
LFI 2015 \rlap{$^{\rm b}$}&              3365.5*\pm*3.0*&     264.01*\pm0.05*&
     48.26*\pm0.02*&\citet{planck2014-a03}\cr
HFI 2015 \rlap{$^{\rm d}$}&              3364.29\pm*1.1*&     263.914\pm0.013&
     48.265\pm0.002&\citet{planck2014-a09}\cr
\noalign{\vskip 3pt}
LFI 2018 \rlap{$^{\rm b}$}&              3364.4*\pm*3.1*&     263.998\pm0.051&
     48.265\pm0.015&\citet{planck2016-l02}\cr
HFI 2018 \rlap{$^{\rm d}$}&              3362.08\pm*0.99&     264.021\pm0.011&
     48.253\pm0.005&\citet{planck2016-l03}\cr
\noalign{\vskip 3pt}
\bf\npipe\ \rlap{$^{\rm a,c}$}& \bf3366.6*\pm*2.7*& \bf263.986\pm0.035&
 \bf48.247\pm0.023&\citet{npipe}\cr
\noalign{\vskip 5pt\hrule\vskip 5pt}}}
\endPlancktablewide
\tablenote {{\rm a}} Statistical and systematic uncertainty estimates
are added in quadrature.\par
\tablenote {{\rm b}} Computed with naive dipole estimator that does
not account for higher-order CMB fluctuations.\par
\tablenote {{\rm c}} Computed with Wiener filter estimator
that estimates, and marginalizes over, higher-order CMB fluctuations jointly with the
dipole.\par
\tablenote {{\rm d}} Higher-order CMB fluctuations are accounted for by subtracting a dipole-adjusted CMB map from frequency maps prior to dipole estimation. \par
\end{table*}

In this paper we have quantitatively compared two numerical techniques
for estimating the dipole parameters from a CMB map. The first method
is basic template fitting regression with partial sky data, while the
second method relies on Wiener filtering. The main difference between
the two methods lies in their treatment of partial sky
observations. Specifically, masking parts of the sky introduces
couplings between small-scale CMB fluctuations and the dipole. The
traditional template fitting procedure disregards this coupling effect,
and simply treats the CMB fluctuations as a noise term. In contrast,
the Wiener filter approach exploits the fact that these fluctuations
represent a statistically isotropic and Gaussian random field to
partially reconstruct the field inside the mask, and thereby reduce
the overall uncertainties. 

We apply both methods to an ensemble of 100 Monte Carlo realizations
for sky fractions ranging from 20 to 95\,\%, and derive uncertainties
as a function of sky fraction. We find that the Wiener filter approach
leads to significantly reduced uncertainties for typical sky fractions
used in this type of analyses. For example, at $f_{\textrm{sky}} \approx60\%$
the uncertainties are reduced by a factor of $\approx3$, while at
$f_{\textrm{sky}} \approx 85\%$ they are reduced by a factor of $\approx8$.

Table~\ref{tab:dipole} shows a comparison of measurements of the CMB
solar dipole made by COBE, WMAP and (various generations of) \Planck,
and is a direct reproduction of Table~10 from \citet{npipe}. Most of
these analyses employed sky fractions around 80\,\%. For this sky
fraction, we see from Fig.~\ref{fig:sigma_plot} that the uncertainty on
the amplitude due to small-scale CMB fluctuations is about $2.5\muK$
using the template fitting approach. In contrast, the total
uncertainty for COBE was 23\muK, and for WMAP it was 8\muK. As such,
the contribution from CMB confusion was sub-dominant for both these
experiments. Nevertheless, it is important to note that WMAP was
indeed the first experiment to implement this method for this
particular purpose \citep{hinshaw2009}, even though it may not have
been critically important.

For \Planck, the situation is fundamentally different. For this
experiment, the raw uncertainty from instrumental noise and
systematics is smaller than 1\muK, and the CMB confusion has therefore
become a dominant factor. In the LFI processing, this contribution was
simply included in the error budget, leading to a final uncertainty of
3\muK. For HFI, however, a different approach was taken, in that an
estimate of the CMB fluctuations was removed from the raw data prior
to template fitting. At first sight, this approach appears to
eliminate the CMB confusion term entirely, evading the topic discussed
in this paper. However, it is important to note that for this approach
to be unbiased, the CMB template that is being subtracted must itself
have a vanishing dipole moment. Determining the dipole moment of this
map is therefore equivalent to the problem described in this
paper. For the HFI analyses summarized in Table~\ref{tab:dipole}, this
determination was performed with a very small mask, which in effect
assumes that the component separation method of choice (see
\citealp{planck2016-l04} for details) is able to remove foregrounds
accurately even in the central galactic plane.

The last row in Table~\ref{tab:dipole} lists results for the most
recent \Planck\ analysis, which is informally referred to as
\npipe. \npipe\ represents the first joint analysis of the
\Planck\ LFI and HFI data sets, using a common machinery to reduce the
raw time-ordered data into final sky maps. One important difference
between these maps and earlier versions of the \Planck\ data is that
the \npipe\ maps retain the solar dipole for both LFI and HFI. It is
therefore, for the first time, possible to compute a single coherent
all-\Planck\ dipole with these sky maps. The results from this
analysis are presented in Sect.~8 of \citet{npipe}, and employ the
Wiener filter methodology described in this paper. The values reported
in Table~\ref{tab:dipole} correspond to a sky fraction of 81\,\%,
which represents a compromise between maximizing available data and
minimizing foreground-induced systematic effects. We believe that the
values reported for \npipe\ in Table~\ref{tab:dipole} represent the
most conservative and statistically robust estimate of the CMB solar
dipole published to date.

\begin{acknowledgements}
  We thank Jeff Jewell and Reijo Keskitalo for useful discussions.
  This work has received funding from the European Union’s Horizon
  2020 research and innovation programme under grant agreement numbers
  776282, 772253 and 819478. Some of the results in this paper have been
  derived using the \healpix\ package.
\end{acknowledgements}

\bibliographystyle{aa}
\bibliography{references}

\end{document}

%% file: Planck.tex
\def\setsymbol#1#2{\expandafter\def\csname #1\endcsname{#2}}
\def\getsymbol#1{\csname #1\endcsname}

\def\Planck{\textit{Planck}}

\newbox\tablebox    \newdimen\tablewidth
\def\leaderfil{\leaders\hbox to 5pt{\hss.\hss}\hfil}
\def\endPlancktablewide{\tablewidth=\textwidth 
    $$\hss\copy\tablebox\hss$$
    \vskip-\lastskip\vskip -2pt}
\def\tablenote#1 #2\par{\begingroup \parindent=0.8em
    \abovedisplayshortskip=0pt\belowdisplayshortskip=0pt
    \noindent
    $$\hss\vbox{\hsize\tablewidth \hangindent=\parindent \hangafter=1 \noindent
    \hbox to \parindent{$^#1$\hss}\strut#2\strut\par}\hss$$
    \endgroup}
\def\doubleline{\vskip 3pt\hrule \vskip 1.5pt \hrule \vskip 5pt}

\def\L2{\ifmmode L_2\else $L_2$\fi}

\def\DeltaT{\ifmmode \Delta T\else $\Delta T$\fi}
\def\deltat{\ifmmode \Delta t\else $\Delta t$\fi}
\def\fknee{\ifmmode f_{\rm knee}\else $f_{\rm knee}$\fi}
\def\Fmax{\ifmmode F_{\rm max}\else $F_{\rm max}$\fi}
\def\solar{\ifmmode{\rm M}_{\mathord\odot}\else${\rm M}_{\mathord\odot}$\fi}
\def\Msolar{\ifmmode{\rm M}_{\mathord\odot}\else${\rm M}_{\mathord\odot}$\fi}
\def\Lsolar{\ifmmode{\rm L}_{\mathord\odot}\else${\rm L}_{\mathord\odot}$\fi}
\def\inv{\ifmmode^{-1}\else$^{-1}$\fi}
\def\mo{\ifmmode^{-1}\else$^{-1}$\fi}
\def\sup#1{\ifmmode ^{\rm #1}\else $^{\rm #1}$\fi}
\def\expo#1{\ifmmode \times 10^{#1}\else $\times 10^{#1}$\fi}
\def\,{\thinspace}
\def\lsim{\mathrel{\raise .4ex\hbox{\rlap{$<$}\lower 1.2ex\hbox{$\sim$}}}}
\def\gsim{\mathrel{\raise .4ex\hbox{\rlap{$>$}\lower 1.2ex\hbox{$\sim$}}}}

\def\simprop{\mathrel{\raise .4ex\hbox{\rlap{$\propto$}\lower 1.2ex\hbox{$\sim$}}}}
\def\deg{\ifmmode^\circ\else$^\circ$\fi}
\def\pdeg{\ifmmode $\setbox0=\hbox{$^{\circ}$}\rlap{\hskip.11\wd0 .}$^{\circ}
          \else \setbox0=\hbox{$^{\circ}$}\rlap{\hskip.11\wd0 .}$^{\circ}$\fi}
\def\arcs{\ifmmode {^{\scriptstyle\prime\prime}}
          \else $^{\scriptstyle\prime\prime}$\fi}
\def\arcm{\ifmmode {^{\scriptstyle\prime}}
          \else $^{\scriptstyle\prime}$\fi}
\newdimen\sa  \newdimen\sb
\def\parcs{\sa=.07em \sb=.03em
     \ifmmode \hbox{\rlap{.}}^{\scriptstyle\prime\kern -\sb\prime}\hbox{\kern -\sa}
     \else \rlap{.}$^{\scriptstyle\prime\kern -\sb\prime}$\kern -\sa\fi}
\def\parcm{\sa=.08em \sb=.03em
     \ifmmode \hbox{\rlap{.}\kern\sa}^{\scriptstyle\prime}\hbox{\kern-\sb}
     \else \rlap{.}\kern\sa$^{\scriptstyle\prime}$\kern-\sb\fi}
\def\ra[#1 #2 #3.#4]{#1\sup{h}#2\sup{m}#3\sup{s}\llap.#4}
\def\dec[#1 #2 #3.#4]{#1\deg#2\arcm#3\arcs\llap.#4}
\def\deco[#1 #2 #3]{#1\deg#2\arcm#3\arcs}
\def\rra[#1 #2]{#1\sup{h}#2\sup{m}}

\def\dots{\relax\ifmmode \ldots\else $\ldots$\fi}
\def\WHzsr{\ifmmode $W\,Hz\mo\,sr\mo$\else W\,Hz\mo\,sr\mo\fi}
\def\mHz{\ifmmode $\,mHz$\else \,mHz\fi}
\def\GHz{\ifmmode $\,GHz$\else \,GHz\fi}
\def\mKs{\ifmmode $\,mK\,s$^{1/2}\else \,mK\,s$^{1/2}$\fi}
\def\muKs{\ifmmode \,\mu$K\,s$^{1/2}\else \,$\mu$K\,s$^{1/2}$\fi}
\def\muKRJs{\ifmmode \,\mu$K$_{\rm RJ}$\,s$^{1/2}\else \,$\mu$K$_{\rm RJ}$\,s$^{1/2}$\fi}
\def\muKHz{\ifmmode \,\mu$K\,Hz$^{-1/2}\else \,$\mu$K\,Hz$^{-1/2}$\fi}
\def\MJysr{\ifmmode \,$MJy\,sr\mo$\else \,MJy\,sr\mo\fi}
\def\MJysrmK{\ifmmode \,$MJy\,sr\mo$\,mK$_{\rm CMB}\mo\else \,MJy\,sr\mo\,mK$_{\rm CMB}\mo$\fi}
\def\microns{\ifmmode \,\mu$m$\else \,$\mu$m\fi}

\def\muK{\ifmmode \,\mu$K$\else \,$\mu$\hbox{K}\fi}
\def\microK{\ifmmode \,\mu$K$\else \,$\mu$\hbox{K}\fi}
\def\muW{\ifmmode \,\mu$W$\else \,$\mu$\hbox{W}\fi}
\def\kms{\ifmmode $\,km\,s$^{-1}\else \,km\,s$^{-1}$\fi}
\def\kmsMpc{\ifmmode $\,\kms\,Mpc\mo$\else \,\kms\,Mpc\mo\fi}
\providecommand{\sorthelp}[1]{}